\documentclass[showpacs,final,a4paper]{appolb}
\usepackage{epsfig}
\usepackage{bm}% bold math
\usepackage{amssymb}
\usepackage{mathrsfs}
\usepackage{amsmath}
\usepackage{dsfont}

\begin{document}
\title{Solving the Complex Phase Problem in a QCD Related Model%
\thanks{Excited QCD - Les Houches 20-25 February, 2011.  Presented by Y.Delgado}%
}
\author{ Ydalia Delgado$^{a}$, Hans Gerd Evertz$^{b}$, Christof Gattringer$^{a}$
%\and
%Hans Gerd Evertz
%\address{\small $^{b}$Institute for Theoretical and Computational Physics, Technische Universit\"at Graz, Austria}
%\and
%Christof Gattringer
\address{\small $^{a}$Institut f\"ur Physik, Karl-Franzens Universit\"at, Graz, Austria}
\address{\small $^{b}$Institute for Theoretical and Computational Physics, Technische Universit\"at Graz, Austria}
}
\maketitle
\begin{abstract}
We discuss an effective theory for QCD at finite chemical potential and non-zero temperature,  where QCD is reduced to its center degrees of freedom.  The effective action can be mapped to a flux representation, where the complex phase problem is solved and the theory accessible to Monte Carlo techniques.  In this work, we use a generalized Prokof'ev-Svistunov worm algorithm to perform the simulations and determine the phase diagram as a function of temperature, quark mass and chemical potential.  It turns out that the transition is qualitatively as expected for QCD.
\end{abstract}
  
\section{Introduction}
Currently one of the main goals of particle physics is to gain a deeper understanding of the QCD phase diagram.  In the next years running and forthcoming heavy-ion colliders will provide access to the regime of QCD phase transitions and challenge theory to describe qualitatively and quantitatively the phase diagram structure.   Due to the non-perturbative nature of QCD, particularly in the transition region, Lattice QCD turns out to be the most suitable tool to describe it.  However, at finite density Lattice QCD faces with the complex phase problem, where the Boltzmann factor becomes complex, making the theory inaccessible to Monte Carlo simulations.  This is the main reason for the very limited progress of Lattice QCD on the description of the QCD phase diagram.  So far the region of small chemical potential is accessible to Lattice QCD \cite{ref1}, while the rest of the phase diagram is still unexplored.  Therefore, it is necessary to develop new ideas and methods, which as a preliminary step can be tested in effective theories.

For quenched QCD the deconfinement transition is related to the center group $\mathds{Z}_3$ of $SU(3)$ \cite{znbreaking}.  When the dynamics of the
quark fields is coupled, center symmetry is broken explicitly, and for $\mu = 0$ it is well established that the transition is a crossover \cite{wuppnature}.
However, one may expect that the underlying symmetry still governs parts of the dynamics of the full theory \cite{canonical}.  
Therefore, in order to study the role of center symmetry for the QCD phase diagram, we analyze an effective theory
which contains the leading center symmetric and center symmetry breaking terms.  This effective center theory can be
mapped exactly to a flux representation \cite{flux}, where the complex phase problem is solved, and which can be simulated using a generalized Prokof'ev-Svistunov
worm algorithm \cite{worm}.  Here we describe the model, the algorithm and the results \cite{dge}.

\section{The Effective Center Theory}
The effective center theory is defined by the action
\begin{equation}
S[P]  = - \!\sum_x \left(\! \tau \! \sum_{\nu = 1}^3 \! \Big[ P_x P_{x+\hat{\nu}}^* + c.c. \Big] 
+ \kappa \Big[ e^\mu P_x +  e^{-\mu} P_x^* \Big]\! \right) ,
\label{action_original}
\end{equation}
where the dynamical degrees of freedom are the elements of the center group, $P_x \in
\mathds{Z}_3  = \{1,e^{i 2 \pi/3}, e^{-i 2\pi/3}\}$ at the sites $x$ of a 3-dimensional lattice.  $P_x$ plays the role of the local Polyakov loop, which in quenched QCD is the order parameter for center symmetry and thus for confinement. The partition function is a sum over all configurations of the center variables, $Z = \sum_{\{P\}}\exp(-S[P])$. 

The nearest neighbor interaction term in (\ref{action_original}), which is invariant under global center transformations, may be obtained from a strong coupling expansion.  The parameter $\tau$ is an increasing function of the temperature $T$ of the underlying Lattice QCD theory.  The second term of the effective action (\ref{action_original}) may be obtained from a hopping expansion of the fermion determinant and contains the leading center symmetry breaking contributions. The parameter $\mu$ is the chemical potential, and $\kappa$ is a decreasing function of the quark mass $m$.

For vanishing $\kappa$, $\ie$ quenched QCD, the model reduces to the 3d 3-state Potts model, which is known to have a first order transition at $\tau = 0.183522(3)$ \cite{potts}. Increasing $\kappa$ at $\mu = 0$ the transition weakens until it terminates in a critical end point at $(\tau,\kappa) = (0.183127(7),0.00026(3))$ \cite{potts}. For small non-zero $\mu$, the chemical potential mildens the transition and shifts the critical endpoint towards smaller values of $\kappa$ \cite{Alford,Kim,Forcrand,Condella}.

\section{Flux representation}
Using high temperature expansion techniques one may map the partition function to a flux representation \cite{flux}.  Its final form is given by:

\begin{equation}
Z \; = \; (3 C^3)^V \sum_{\{b,s\}}  
\left( \prod_{x,\nu} B^{|b_{x,\nu}|} \right) \left(
\prod_x \, M_{s_x} \right) \, \prod_x  T \left( \sum_\nu [b_{x,\nu} - b_{x-\hat{\nu},\nu}] 
+ s_x \right) \;
\label{zflux}
\end{equation}

Here, $C =(e^{2\tau} + 2 e^{-\tau})/3$, $B = \; (e^{2\tau} - e^{-\tau})/3C$, and the monomer weights $M_s$ for $s = -1,0,+1$ are:

\begin{equation}\!
M_s\; =\; \frac{1}{3} \! \left[ e^{\,2\kappa\cosh\mu} \, + \, 
2 \, e^{-\kappa\cosh\mu} \, \cos \left( \sqrt{3}\kappa\sinh\mu - \frac{2\pi}{3}s) \right) \right]
\label{monoweightn}
\end{equation}

In the new representation (\ref{zflux}) the configurations are closed paths formed 
by fluxes, $b_{x,\nu} \in \{-1,0,+1\}$ which live on the
links $(x,\nu)$, and monomers $s_{x} \in \{-1,0,+1\}$ attached to the sites $x$. They
must obey the constraint given by the second factor in (\ref{zflux}). The triality function $T(n)=\delta_{0,n\ mod\ 3}$ 
enforces the total sum of fluxes and monomers to be a multiple of 3 at each site $x$.
Each configuration of dimers and monomers comes with a strictly positive weight and thus the complex phase problem of this model is solved.

We focus on bulk observables such as the expectation value of the Polyakov loop $P$, its corresponding susceptibility $\chi_P$, the internal energy $U$ and the heat capacity $C$, which all are obtained as derivatives of the free energy. In the end all our observables are expressed in terms of the total number of fluxes and monomers.
      
\section{Simulation with the worm algorithm}
\begin{figure}[t]
\begin{center}
\includegraphics[width=10cm,clip]{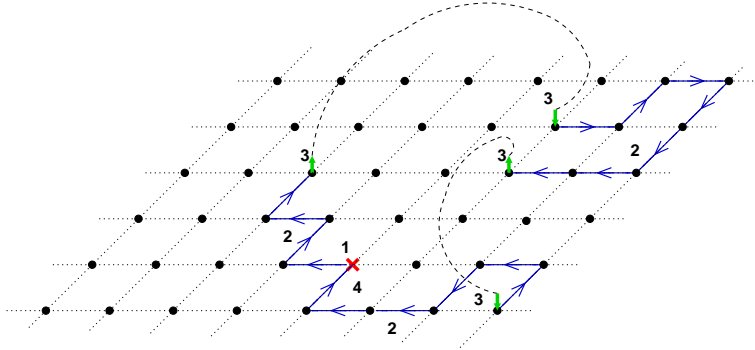}
\end{center}
\vspace{-4mm}
\caption{Schematic illustration of the worm algorithm.\vspace{-4mm}}
\label{worm1}
\end{figure}

The most suitable algorithm to perform the Monte Carlo simulation of our model is a generalized form of the Prokof'ev-Svistunov worm algorithm \cite{worm}.  Each worm configuration is generated using four different steps (Fig.~\ref{worm1}).  The worm starts at a random position of the lattice (1).  It may decide to insert fluxes and move to the neighboring site (2) but can also insert monomers (3).  The insertion of a monomer is followed by a random hop to another position (3), where again a monomer is  inserted.  These steps are continued until the worm closes (4).

The results from the new algorithm were checked using several strategies.  For the quenched case ($\kappa = 0$) the known results were reproduced \cite{potts}. For small values of $\tau$ we determined the power series perturbatively for the partition sum, taking into account the terms up to $\tau^3$.  We found excellent
agreement between the Monte Carlo results and the perturbative series (see, Fig.~\ref{phasediagram_chiP}). Finally, two independent programs were written for cross checks.

We generated ensembles on  $36^3$, $48^3$, $64^3$ and $72^3$ lattices, for $\kappa = 0.1,\ 0.01,\ 0.005$ and $0.001$. The evaluation of our observables $\langle P \rangle$, $\chi_P$, $U$ and $C$ is based on up to $10^6$ configurations which are separated by 10 worms for decorrelation. Autocorrelation times were determined for several observables and used in the estimate for the statistical errors.

\section{Results from the Monte Carlo simulation}
\begin{figure}[t!]
\begin{center}
\includegraphics[width=5.8cm,clip]{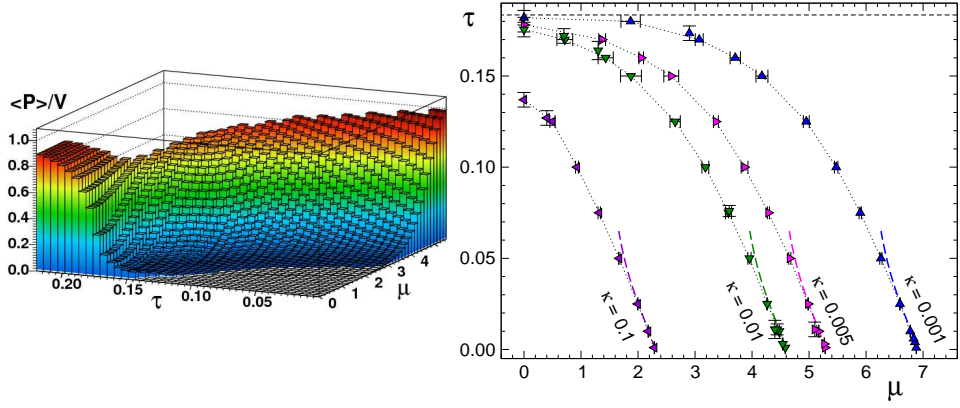}
\includegraphics[width=6.6cm,clip]{pics/phaseline_Xp.eps} 
\end{center}
\vspace{-4mm}
\caption{{\it Left:} Order parameter $\langle P\rangle/V$ as a function of $\tau$ and $\mu$ for $V=36^3$ at $\kappa = 0.01$. {\it Right:} Phase diagram obtained from the maxima of the Polyakov loop susceptibility for 4 values of $\kappa$.  The dashed horizontal line marks the critical value of $\tau$ for $\kappa = 0$, and the dashed curves at the bottom are the results from the perturbative $\tau$ expansion.\vspace{-4mm}}
\label{phasediagram_chiP}
\end{figure}

The lhs. plot of Fig.~\ref{phasediagram_chiP} shows $\langle P\rangle/V$ as a function of $\tau$ and $\mu$ at $\kappa = 0.01$.  For small $\tau$ and $\mu$, $\langle P\rangle/V$ is rather small and center symmetry is broken only very mildly, which implies that matter is confined.  On the other hand, when $\tau$ and $\mu$ increase, the system undergoes a change and $\langle P\rangle/V$ reaches values close to $1$.  For QCD this means that the system is driven into the deconfined phase signaled by a rapid increase of the Polyakov loop.

To identify the phase boundary in the $\tau$-$\mu$ plane (Fig.~\ref{phasediagram_chiP} rhs. plot), we use the position of the maxima of $\chi_P$.  For this determination we fit the data for $\chi_p$ near the maxima with a parabola and obtained the position of the maximum as one of the fit parameters.  The corresponding statistical error was computed with the jackknife method.  The points with horizontal (vertical) errors were determined at fix $\tau$ ($\mu$) as a function of $\mu$ ($\tau$). We compare the results for 4 values of $\kappa$. The dashed horizontal line at the top marks the value of the critical $\tau$ for $\kappa = 0$.  The dashed curves near the bottom of the plot are the results from the perturbative series for small $\tau$, which we briefly discussed in the last section.

\begin{figure}[t!]
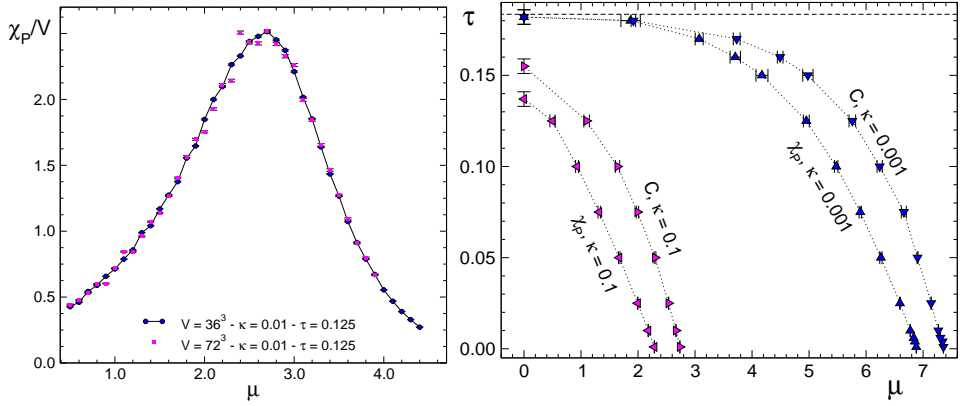

\begin{center}
\includegraphics[width=5.8cm,clip]{pics/maxima_xp.eps}
\includegraphics[width=6.6cm,clip]{pics/phaseline_compare.eps}
\end{center}
\vspace{-4mm}
\caption{{\it Left:} The critical parameters are determined by locating the position of the maxima of the susceptibility $\chi_P$. {\it Right:} Comparison of the phase boundaries obtained from the maxima of susceptibility $\chi_P$ and heat capacity $C$. \vspace{-4mm}}
\label{phasediagram_chiP_C}
\end{figure}

To determine the nature of the phase transition we compared the phase boundaries of the two order parameters and the heights of their fluctuations at different volumes.  We found that the transition lines from $\chi_P$ and $C$ do not coincide (Fig.~\ref{phasediagram_chiP_C} rhs. plot), and there is no volume scaling (Fig.~\ref{phasediagram_chiP_C} lhs. plot).  This is a clear signal that the phase changes are crossovers.  Although there is no critical end point when only center degrees of freedom are considered, the qualitative behavior and the mass dependence of the phase boundaries are as one expects for full QCD.

\section{Conclusions}
We have studied an effective theory of QCD with finite chemical potential at non zero temperature. 
Mapping this theory to the flux representation enables us not only to have a model free of the complex phase 
problem but also to generate configurations efficiently in a wide range of parameters using a worm algorithm.  
For all parameter values we studied the transition is of a crossover type and we conclude that center symmetry alone
does not provide a mechanism for first order behavior in the QCD phase diagram.

Currently we consider two possible further directions: The effective theory can be made more realistic by replacing the $\mathds{Z}_3$ spins by continuous $SU(3)$
valued variables \cite{cg}.  Also, we have implemented a parallel version of the worm algorithm, to take advantage of the compute capability of graphic cards.
\\
\\
{\bf Acknowledgments: } The authors thank Philippe de Forcrand, Christian Lang and Bernd-Jochen Schaefer for valuable discussions and remarks. The project is supported by FWF DK 1203 and Marie Curie ITN STRONGnet.

\end{document}